\documentclass{article}

\usepackage[
    backend=biber,
    style=authoryear,
    maxcitenames=2,      
    maxbibnames=99,      
    uniquelist=false,
    giveninits=true,     
    terseinits=true,     
    dashed=false,        
    isbn=false,          
    doi=true,            
    uniquename=false,
    url = false,
    eprint = false]{biblatex}
\AtEveryBibitem{\clearfield{note}}
\renewbibmacro{in:}{}

\usepackage[english]{babel}

\usepackage[letterpaper,top=2cm,bottom=2cm,left=3cm,right=3cm,marginparwidth=1.75cm]{geometry}
\usepackage{parskip}

\usepackage{amsmath}
\usepackage{graphicx}
\usepackage[colorlinks=true, allcolors=blue]{hyperref}

\usepackage{authblk}
\usepackage{float}

\title{Temporal Structure Mediates the Robustness and Collapse of Plant–Pollinator Networks}

\author[a,b]{Tom Clegg}
\author[a,b,c]{Thilo Gross}

\affil[a]{Helmholtz Institute for Functional Marine Biodiversity at the University of Oldenburg, Im Technologiepark 5, 26129 Oldenburg, Germany.}
\affil[b]{Alfred Wegner institute, Am Handelshafen 12, 27570 Bremerhaven, Germany}
\affil[c]{Institute for Chemistry and Biology of the Marine Environment, Carl-von-Ossietzky-Straße 9-11, 26111 Oldenburg, Germany}
\date{}    

\begin{document}
\maketitle

\subsubsection*{Conflict of Interest}
The authors declare no conflicts of interest.

\subsubsection*{Author Contributions}
All authors contributed to conception of the study, analysis of the model and drafting and revising the manuscript.

\subsubsection*{Data availability statement}
All code used to generate results and figures is available at \href{https://doi.org/10.5281/zenodo.19455984}{doi.org/10.5281/zenodo.19455984}  

\newpage
\begin{abstract}
Mutualistic networks provide a powerful way to describe and analyse plant–pollinator communities and their structure over time. While these networks capture the complex interdependencies that link population fates across the season, they can be hard to untangle, preventing us from understanding the emergence of community-scale properties and responses to perturbation. Here, we address this problem by developing a structural model of a plant–pollinator community that explicitly incorporates seasonal turnover and the temporal nature of species interactions. We analyse our model using percolation methods from network science to derive simple analytical solutions linking network structure to emergent community diversity. Our findings reveal that temporal structure organises community diversity into distinct ecological phases, creating the potential for alternative high- and low-diversity states and bistable regimes. We demonstrate how this temporal structure mediates the nature of transitions between these states, determining whether systems undergo gradual shifts or abrupt, catastrophic collapses. Crucially, we show how this temporal structure reduces the robustness of plant–pollinator systems, creating bottlenecks that inhibit species persistence and increase susceptibility to secondary extinctions. Our results demonstrate that the temporal dynamics of plant–pollinator networks are central to mediating their fragility, highlighting the importance of accounting for time when considering community resilience.
\end{abstract}

{\bf Keywords:}  Plant–Pollinator Networks, Temporal Networks, Robustness, Percolation Theory

\section*{Introduction}
Mutualistic interactions play a fundamental role in plant–pollinator communities, linking populations through the exchange of floral resources and pollination services \parencite{bascompte_plant-animal_2007}. Ecologists frequently represent these interactions as a bipartite network, in which plants and animals form two sets of nodes connected by mutualistic links. This approach provides a powerful way to describe and analyse the structure of plant–pollinator communities, and has yielded insight into the emergence of key properties such as community functioning \parencite{timberlake_network_2022}, coexistence \parencite{bastolla_architecture_2009}, and robustness to perturbation \parencite{memmott_tolerance_2004}.

One of the major challenges in studying plant–pollinator networks is accounting for their temporal structure. Plant–pollinator communities are inherently dynamic, driven by the species turnover that arises from life-history events such as flowering, pollinator emergence, and senescence \parencite{caradonna_seeing_2021, ogilvie_interactions_2017}. The timing of these events is governed by phenology \parencite{rathcke_phenological_1985}, which imposes constraints on species interactions \parencite{olesen_missing_2010} but is increasingly being disrupted by anthropogenic pressures such as climate and land-use changes \parencite{kudo_when_2019}. Consequently, the organisation and resilience of these communities are poorly captured by static networks and instead require a representation that explicitly includes temporal structure \parencite{caradonna_seeing_2021}.

One approach that is often used to capture this structure is to represent communities as a series of snapshots across time. By aggregating species and their interactions into discrete windows, studies have been able to characterise how network topology and function change over seasons \parencite{petanidou_long-term_2008, trojelsgaard_ecological_2016}, and attribute these changes to processes such as species turnover and interaction rewiring \parencite{caradonna_interaction_2017, caradonna_temporal_2020, lampo_structural_2024}. While this approach successfully captures the shifting topology of plant–pollinator networks, it does not account for the coupling of populations across time: a species' persistence in one snapshot depends on the conditions it faces across other parts of the season. 

This temporal coupling arises, in part, from species' ecological demands. For example, many active pollinators require constant access to floral resources throughout their life cycles to fuel maintenance and reproduction \parencite{russo_supporting_2013, schellhorn_time_2015, nicholson_corridors_2021, lowe_temporal_2023}. Conversely, plants only need pollination during discrete flowering periods to facilitate reproduction and, ultimately, population persistence \parencite{knight_pollen_2005}. Together, these requirements link the viability of a population to the success of its partners across their life cycles, who in turn rely on the survival of their own partners. As a consequence, the persistence of any individual species becomes dependent on a chain of interactions across time, which at a system-level form a network of temporal interdependencies. 

While the structure of these interdependencies is important, its complexity poses a significant challenge. The arrangement of these links influences community-level properties like diversity and can facilitate abrupt, system-wide collapse \parencite{lever_sudden_2014}, a risk of increasing concern in the context of the Anthropocene \parencite{bascompte_resilience_2023}. Characterising emergent phenomena like these collapses is non-trivial because of the combinatorial complexity of enumerating dependencies between species, which can quickly become computationally infeasible in larger systems. Furthermore, because the effect of interactions relies heavily on their local configuration, it remains difficult to identify general patterns that can be applied across different network structures.

The problem of how the structure interdependencies determine system-level properties is not unique to plant–pollinator communities. This phenomenon has long been studied through the lens of network percolation theory, a branch of network science that examines how local dependencies influence system-wide connectivity and function \parencite{callaway_network_2000,newman_random_2001,buldyrev_catastrophic_2010}. Translating this framework to ecology, recent work on microbial communities framed the emergence of community-scale properties in terms of the appearance of a giant connected component, a self-sustaining core of populations which can mutually facilitate one another's survival \parencite{clegg_cross-feeding_2025}. This approach allows the use of tools such as generating functions to study not only how network architecture governs community diversity and robustness, but also how it dictates the emergence of alternative community states and the nature of transitions between them \parencite{newman_random_2001, gross_network_2022}. Furthermore, these methods enable us to consider whole ensembles of networks that satisfy a set of structural constraints simultaneously, letting us derive generalisable insights that apply across a broad range of network structures. 

In this study, we investigate how temporal network structure determines the organisation of plant–pollinator communities and their response to disturbance. By developing a general model of seasonally driven plant–pollinator networks, which we analyse through the lens of percolation theory, we explicitly account for the interdependencies created by population persistence across time. We demonstrate how the statistical properties of the temporal network structure act as drivers of community diversity, organising systems into distinct high- and low-diversity states. Furthermore, we characterise the transitions between these alternative states, showing how the inclusion of temporal dependencies can render communities more fragile, and facilitate abrupt, catastrophic collapses. Finally, we show that temporal structure has profound consequences for community robustness, highlighting the need to account for temporal coupling when assessing the resilience of plant–pollinator communities.


\section*{The Seasonal Plant–pollinator Model}
\begin{figure}
    \centering
    \includegraphics[width=0.9\linewidth]{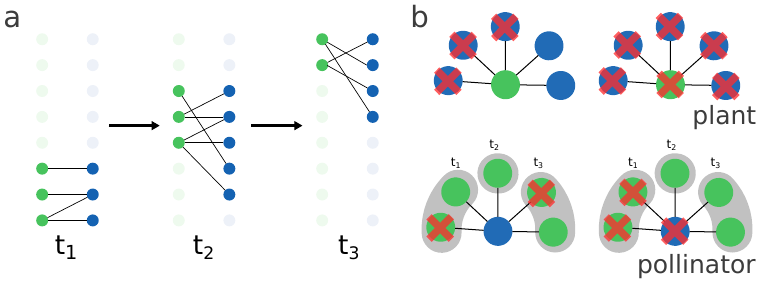}
    \caption{{\bf The temporal plant–pollinator network model.} a) Temporal turnover in plant (green) and pollinator (blue) activity of nodes results in changes in network topology. Each network shows a the bipartite network structure in a given time step as nodes switch between active (coloured) and inactive (greyed) states. As time progresses links form between active nodes and are lost as others become inactive. b) The feasibility of nodes is determined by simple rules which depend on the state of their neighbours across time. Plants need at least one feasible pollinator neighbour (top) whilst pollinators need a feasible neighbour at every time step (each collection of nodes) (bottom).}
    \label{fig:Diagram}
\end{figure}

We model the seasonal dynamics of a mutualistic community comprising $N$ plant and $M$ pollinator species. The community is represented as a bipartite network, in which nodes are species populations linked by their mutualistic interactions. This network captures the global structure of community, and is made up the set of potential interactions between species. The model progresses through time in discrete steps, capturing species turnover across the season (Fig~\ref{fig:Diagram}a). Plant nodes are active for a single step, representing a short flowering period, while pollinator nodes emerge and remain active for a multiple steps, capturing the longer timescales required to support their lifecycles. 

Within this framework, we consider a species viable if it is able to re-establish after a local extinction due to external events. Hence, two species locked in obligate mutualism would not be considered viable, because the extinction of one of them would lead to the loss of the other, precluding either of them from re-establishing on their own.  

To assess viability, we define simple rules that capture the distinct ecological dependencies of plant and pollinator species (Fig~\ref{fig:Diagram}b). Plants are able to persist if they interact with at least one viable pollinator, assuming that pollination by a single species is sufficient to allow reproduction and thus population persistence. Pollinators face stricter requirements, persisting only if they have access to at least one viable plant at every step of their active period. This rule reflects pollinators' continuous need for food resources, such as nectar and pollen, to fuel their metabolism, growth and reproduction \parencite{schellhorn_time_2015,nicholson_corridors_2021}.

To move beyond the specific structure of any single interaction network and gain a broader understanding of community persistence, we study our model using a statistical ensemble approach. Rather than focusing on a fixed set of links, we analyse the behaviour of the model across the set of all possible networks that satisfy a set of structural constrains. We define this ensemble through the degree distributions within each step for plants $p_k$ and pollinators $a_k$  and the length of pollinator active periods $l_t$. In this notation, $p_3$ and $a_3$ are the probabilities that a random plant or pollinator has three links in a given step while $l_2$ would be the probability a pollinator is active for two steps. Together, these distributions define the network topology: the node degrees control the connectivity within each step, while the active period length governs the linkages that define the extent of each pollinator's influence across time.

Across the ensemble of network structures, we focus on calculating the proportion of plant ($p$) and pollinator ($a$) species that meet the feasibility conditions outlined above. While these values represent the probability that any individual species is viable across the ensemble, they also define the realised diversity of the community, measuring the relative size of its self-sustaining core. Our goal is to determine how these values emerge from underlying network structure, identify different states of community organisation and characterise potential transitions between them.

\subsection*{Generating Function Analysis}
While the local rules for species viability are straightforward, calculating the emergent diversity of plants and pollinators across the ensemble of network structures is a non-trivial task. To address this, we use a formalism based on mathematical objects called generating functions to encode and manipulate sequences of numbers \parencite{newman_random_2001, gross_network_2022,clegg_cross-feeding_2025}. This framework allows us to map local ecological dependencies onto a system of equations describing community-wide persistence. To capture the structural properties of the network ensemble, we use generating functions to represent the degree ($p_k$, $a_k$) and pollinator active period ($l_t$) distributions. Considering the plant degree generating function first as an example, we define
\begin{align}
    P(x) = \sum p_k x^k = p_0 + p_1 x + p_2x^2+\dots \label{eq:plant_GF}
\end{align}
where each coefficient of $x^k$ corresponds to the probability that a plant interacts with $k$ pollinators. We similarly define $A(x)$ and $L(x)$ to encode the pollinator degree and active period length distributions respectively
\begin{align}
    A(x) = \sum a_k x^k \quad \quad L(x) = \sum l_{t} x^t \label{eq:pol_GF}
\end{align}

\subsubsection*{Community-Level viability, ($p,a$)}
To characterise community-level diversity across the ensemble, we will first derive equations for the proportion of viable plant ($p$) and pollinator ($a$) populations. This is equivalent to the probability that a typical (i.e.,~randomly chosen) plant or pollinator population is viable (Fig~\ref{fig:excess}). 

According to the criteria for persistence, a plant is viable if it has at least one neighbouring pollinator that can survive in the system \emph{independently} of the plant itself. This independence ensures the plant can re-establish following local extinction. It follows that a plant is not viable if all its potential pollinators fail to persist in its absence. 

Let's now suppose that each of the potential pollinator neighbours of our plant, defined as any species with which it shares a structural link, is viable with probability $b$. Conversely, a neighbour is non-viable with probability $1-b$. By assuming the system is large and possesses locally tree-like structure (i.e. short loops are rare) we can treat these neighbour states as independent. Under this approximation, the probability that none of a plant's $k$ neighbours are able to persist is $(1-b)^k$. To compute the probability that a typical plant is non-viable (i.e.~probability $1-p$) we average this quantity over the plant degree distribution, $p_k$, which yields 
\begin{align}
    1-p = \sum p_k (1-b)^k = P(1-b), 
\end{align}
where we use the definition of the generating function $P(x)$ in the last step. Rearranging for $p$, we obtain the final expression for the fraction of feasible plant populations
\begin{align}
    p = 1 - P(1-b), \label{eq:p}
\end{align}
which expresses the probability a plant can persist as a function of viability of its neighbouring pollinators.

We can apply similar logic to determine the fraction of feasible pollinator species ($a$), though their temporal requirements introduce an extra layer of complexity. Let's first consider the probability $s$ that a pollinator's needs are met in a single step. This is determined by the availability of at least one viable plant neighbour \emph{independently} of the focal pollinator's state in that step. Letting $q$ represent the probability that a neighbouring plant is viable, we compute $s$ by averaging the probability of having at least one partner across the degree distribution $a_k$
\begin{align}
    s = 1-\sum a_k(1-q)^k = 1-A(1-q).
\end{align}

Across time, pollinators face a second condition for feasibility: they must have access to a viable food source in every step of their active period. Since each plant flowers for a single step, the set of interactions for each pollinator is effectively redrawn at every step. This turnover ensures the floral resources available in successive intervals are independent, meaning that the probability a pollinator remains viable across $t$ steps is the product of its individual successes, $s^t$. Averaging this quantity over the over the active period distribution ($l_t$) yields the expression for pollinator viability
\begin{align}
    a = \sum_t l_t s^t = L(s) \label{eq:a}
\end{align}
Together with Equation~\ref{eq:p}, we can expand the definition of $s$ to get the coupled system which defines the state of the plant–pollinator community
\begin{align}
    \begin{split}
        p &= 1-P(1-b) \\
        a &= L(1 - A(1-q),\label{eq:sys}
    \end{split}
\end{align}
showing how fraction of feasible populations is linked to the viability of their partners ($q,b$), mediated by the structure of the network via the generating functions $P(x),A(x)$ and $L(x)$.

\subsubsection*{Neighbour Viability, ($q,b$)}
\begin{figure}[H]
    \centering
    \includegraphics[width=0.5\linewidth]{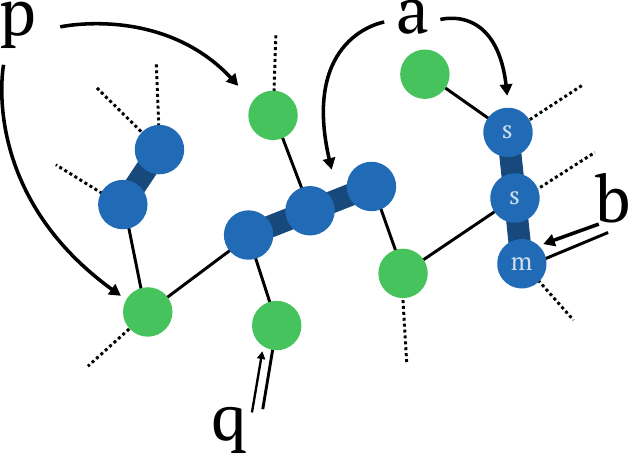}
    \caption{{\bf Network diagram illustrating the calculation of system diversity ($p,a$) and neighbour viability ($q,b$).} The diagram shows a subset of a larger community in which plants species are represented by green nodes and pollinators by a set of blue nodes, Each node in the pollinator set represents a single step of their active period. Solid lines indicate focal interactions, while dashed lines represent links to the rest of the community. The probabilities $p$ and $a$ characterise the viability of a typical plant or pollinator, corresponding to a random sampling of nodes across the system. The probabilities $q$ and $b$ represent neighbour viability, calculated by following a link to a node and then determining its feasibility based on the additional interactions it takes part in. The diagram demonstrates how this applies to the temporal requirements for pollinators: we arrive via a link to a specific step in its active period (which is viable with probability $m$) and then account for persistence across the other active intervals (each with probability $s$).}
    \label{fig:excess}
\end{figure}

The system defined in Equation~\ref{eq:sys} provides a way to compute the fraction of feasible plants and pollinators ($p,a$) based on the probability that their neighbours are viable ($q,b$). Although we could approximate these neighbour probabilities by treating them as equivalent to a random sampling of the community (i.e., $p \approx q$ and $b \approx a$), network science  tells us that it is valuable to be a little more careful \parencite{newman_random_2001}. Because the neighbouring nodes represented in $q$ and $b$ are reached via an existing interaction, their sampling is biased towards high-degree nodes, meaning they are disproportionately likely to interact with many other species. Moreover, because we are interested in the probability that these neighbours are viable \emph{independently} of the focal node, the task of calculating $q$ and $b$ shifts to determining whether a neighbour's additional interactions are sufficient to facilitate its own persistence (Fig~\ref{fig:excess}).

We compute these quantities by first considering the probability that a plant neighbour is viable independently of the focal pollinator with which it interacts, $q$. We will break this calculation into two steps, determining the how many other pollinators the plant interacts with, and then the probability it is viable based on these interactions. 

To determine the number of additional interactions, we need to perform a calculation that is analogous to the determination of excess degree in network science \parencite{newman_random_2001}. We start by defining $q_k$ as the probability that a neighbouring plant interacts with $k$ \emph{additional pollinators}. Clearly, this is identical to the probability that the plant interacts with $k+1$ pollinators in total, as the interaction we used to reach the plant must be excluded from the count of its additional partners. To compute $q_k$, we need to determine the proportion of all links that lead from pollinators to plants of degree $k+1$,
\begin{align}
    q_k = \frac{\mbox{Number of links to plants with $k+1$ interactions}}{\mbox{Number of all links to plants}}
\end{align}

The number of all links that lead to plants is the product of the mean plant degree, $z$, and the total number of plant nodes, $N$
\begin{align}
   N z =  N \sum p_k k = N P'(1)      
\end{align}
where $P'$ is the derivative of the generating function for plant degree. Similarly, the number of links leading to plants with $k+1$ pollinators is $N(k+1)p_{k+1}$, as each of the $N$ plant species interacts with $k+1$ pollinators with probability $p_{k+1}$. Putting these expressions together, the $N$ factors cancel, yielding  
\begin{align}
    q_k = \frac{p_{k+1}(k+1)}{P'(1)}. \label{eq:qk}
\end{align}
This lets us define the generating function that encodes the distribution of additional interactions $Q(x)$, which can be written in terms of the normal plant degree generating function $P(x)$
\begin{align}
    Q(x) = \sum q_k x^k = \sum \frac{q_{k+1} (k+1)}{P'(1)} x^k =  \frac{1}{P'(1)}\sum p_k k x^{k-1} = \frac{P'(x)}{P'(1)},  
\end{align}
where we use the definition of $q_k$ in equation~\ref{eq:qk} and shift the index of the summation in the third step. 

With the number of additional interacting partners captured in $Q(x)$, we can now compute the probability that the plant neighbour is viable based on these other pollinators. This follows the same logic as the expression for $p$ in Equation~\ref{eq:p}, though this time we ask what is the probability that at least one of the \emph{additional} $k$ pollinator neighbours is active independently of the plant under consideration. Averaging this quantity over $q_k$ yields
\begin{align}
    q = 1-\sum_k q_k (1-b)^k = 1 - Q(1-b), \label{eq:q}
\end{align}
which computes the probability a random plant neighbour is viable ($q$) in terms of the state of its additional pollinator neighbours ($b$).

The corresponding equation for the neighbouring pollinators of a focal plant ($b$) follows the similar logic, but requires extra care when considering their more complex temporal requirements. We start by defining the generating functions for the number of additional plants that interact with these pollinators in each time step, $B(x) = \sum b_k x^k$ and the analogous remaining active period distribution $K(x) = \sum k_t x^t$
\begin{align}
    B(x)= \frac{A'(x)}{A'(1)} \quad K(x) = \frac{L'(x)}{L'(x)}
\end{align}
The distribution of additional interactions ($b_k$) has a similar interpretation as $q_k$, capturing the number of interacting partners a neighbouring pollinator has in a single step, excluding the plant from which we encounter the pollinator. The remaining active period distribution $k_t$ captures the number of additional steps that a pollinator is active for, given it was encountered by a flowering plant in a given step. Much like the bias towards high-degree nodes in $q_k$ and $b_k$, this temporal sampling disproportionately favours pollinators with longer active periods as they are more likely to be chosen when we sample a random point in time.

To compute the total probability a pollinator neighbour is viable ($b$) we need to consider the probability their needs are met in each of the steps of its active period, though we need to be careful about how we count the other plants it interacts with and, importantly, when these interactions occur. To facilitate this, we can break this calculation into the product of viability in two more manageable parts: 1) the step in which we encounter the pollinator and 2) the remaining steps of its active period.

To address the first part, we let $m$ be the probability that a neighbouring pollinator's needs are satisfied in the step it was encountered. For a pollinator that interacts with $k$ additional plants, the probability that none are viable is $(1-q)^k$. Averaging over the distribution of additional interactions, $b_k$, and solving for $m$ yields the expression for viability in the encountered time step
\begin{align}
    m = 1 - \sum b_k (1-q)^k = 1 - B(1-q),
\end{align}
which is similar to the expression for needs being met in a single step ($s$), but counts only the additional plant neighbours using $B(x)$, thus ensuring independence from the focal plant node.  

Next, we consider how the needs of the pollinator are satisfied over the remaining $t$ time steps in its active period. Because each of the additional time steps are independent of the step we encounter the pollinator, we can use the product of per-step successes to calculate the probability that the pollinator's needs are met, $s^t$. Averaging over the distribution of remaining time steps then yields the total probability a pollinator neighbour meets its viability condition over all remaining steps $\sum k_t s^t = K(s)$.

Putting these two parts together, we obtain the final expression for the probability that a pollinator neighbour is active
\begin{align}
    b = m K(s), \label{eq:b}
\end{align}
which accounts for the step we encounter the pollinator in $m$, and then the remainder of its active period in $K(s)$.

Combining Equations~\ref{eq:q}~and~\ref{eq:b} and expanding the definitions of $m$ and $s$ yields the final closed system of equations that describe the neighbour states 
\begin{align}
\begin{split}
    q &= 1 - Q(1 - b) \\
    b &= [1 - B(1-q)]K(1 - A(1-q)) \label{eq:sys_neighbour}
\end{split}
\end{align}
Once a network structure is defined by selecting appropriate distributions for $P(x)$, $A(x)$, and $L(x)$ we can use these equations to find solutions for $q$ and $b$, which in turn can be used to solve for the realised diversity in the system $p$ and $a$.

\subsection*{Random Plant Pollinator Networks}
To explore the model's behaviour and the relationship between diversity and community structure, we consider the case of random plant–pollinator networks based on Erdős–Rényi networks \parencite{erdos_random_1959} used across network science and ecology. We focus on this general case to build intuition about the behaviour of the model and because the simple functional forms make parameterisation and interpretation easy. 

We construct random plant–pollinator networks by considering a community in which species are randomly distributed across time and each species pair is allowed to interact with a fixed probability if their active periods overlap. We use a shifted Poisson distribution for the pollinator active period length ($l_t$), so that all pollinators are active for at least one step. This is controlled by the mean active period length $\tau$. Allowing interactions to form with a fixed probability results in a Poisson degree distribution for plant ($p_k$) and pollinator ($a_k$) node degrees, both of which are parameterised by the mean degree per step $z$. Together, $z$ and $\tau$ govern the network structure, determining the connectance within and across time steps, respectively. The Poisson distribution generating function has a simple exponential form $F(x) = e^{z(x-1)}$, which can be substituted directly into equations~\ref{eq:sys} and \ref{eq:a} to obtain solutions for $p$ and $a$.

We also contrast our analytical results with numerical simulations to validate our analytical predictions against randomly generated fixed network structures. For a given set of parameters ($z,\tau$), community size ($N,M$) and number of time steps $T$ we generate a random network, drawing node degrees and active period lengths from the prescribed distributions. We then find feasible community states by initialising the system with a small number of active species and then iterating through each node and checking its viability individually. This process is repeated until the reaches a steady state, after which we calculate the final proportion of viable plants and pollinators. We repeat this across many network realisations to compare the average result with the analytical solutions from equations~\ref{eq:sys}.

\subsection*{Network structure determines patterns of community diversity}
\begin{figure}[H]
    \centering
    \includegraphics[width=0.75\linewidth]{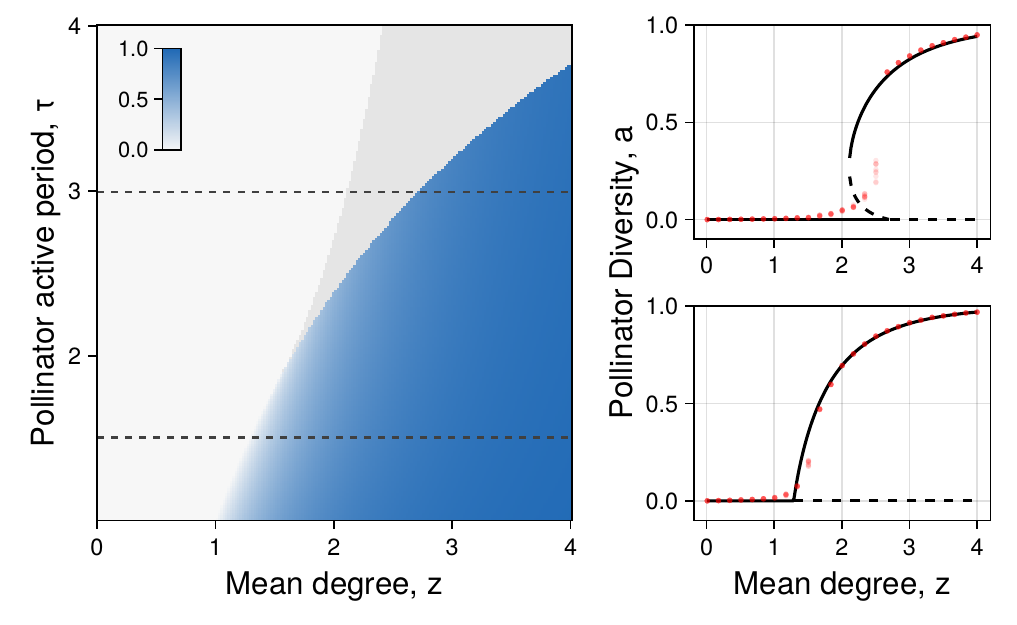}
    \caption{\textbf{Network structure drives changes in pollinator diversity}, a) Phase plot showing how pollinator diversity changes with network structure, shifting from a zero (white) to non-zero (blue) diversity. The change is initially continuous, but pollinators temporal requirements create discontinuous transitions around the bistable region (grey). b \& c) Slices along the phase plot (dashed lines) show the change in the nature of the transition, from continuous (b) to discontinuous (c). Lines show the analytical predictions from equation~\ref{eq:sys} and points the numerical simulations with matching distributions and $N \approx10^5$ and $T=10$.}
    \label{fig:Fig1}
\end{figure}

Solving for the relative diversity of pollinators $a$ in the random networks shows how diversity increases with the mean degree $z$ and decreases with the mean pollinator active period length $\tau$ (Fig~\ref{fig:Fig1}). This pattern is driven by the fact that longer active periods make pollinator requirements harder to satisfy, increasing the frequency of temporal bottlenecks, time steps in which pollinators' needs are not met. This reduces the chance that pollinators and the plants they support are able to persist, reducing the size of the feasible community. Conversely, increasing the plant and pollinator mean degree $z$ increases the chance that either can interact with viable populations within each step, enhancing the likelihood their respective needs are met, allowing greater diversity to persist.   

The solutions to equations~\ref{eq:p} and \ref{eq:a} demonstrate how distinct phases of community organisation arise from specific network structures (Fig~\ref{fig:Fig1}a). When pollinator active periods $\tau$ are short and network connectivity is low, $z < 1$, the system is trapped in a zero-diversity state; the sparsity of interactions means that the chance that any species' needs are met is very low. As connectivity increases $z$, diversity remains at zero until a critical threshold is reached, at which point a non-zero feasible state appears (Fig~\ref{fig:Fig1}c). In the regime of short $\tau$, this transition is continuous but abrupt, representing a gradual growth of the feasible community, and is identical to the classic giant component transition seen in random networks \parencite{molloy_critical_1995}. 

As pollinator active periods get longer, the network becomes more temporally connected, and the transition gets sharper. Once $\tau$ passes a critical threshold, the nature of the transition changes and becomes discontinuous (Fig~\ref{fig:Fig1}b). In this regime, as the system approaches the critical point even small changes in the structure of the interaction network can trigger sudden jumps between feasible and zero-diversity states. This region also marks the appearance of bistability, where both the zero-diversity and feasible states can coexist under the same structural conditions (grey area in Fig~\ref{fig:Fig1}a). 

This phenomenon is driven by the asymmetric requirements of plant and pollinator species. As $\tau$ increases, it becomes harder to satisfy the pollinators' continuos resource requirements, keeping the system in the zero-diversity state even as the network becomes more connected. When $\tau$ is high this creates a build-up of ``unused support'': plants are structurally capable of persisting, but cannot because the pollinators they depend on are still suppressed by temporal bottlenecks.

As connectivity increases, the system eventually reaches a critical threshold where the network of interactions is dense enough to overcome these bottlenecks. At this point, a self-reinforcing feedback loop forms: the activation of a few plants enables more pollinators to persist which, in turn facilitates other plants, causing community diversity jump into a feasible state. The same mechanism operates in reverse as $z$ is reduced in the feasible state; active plants and pollinators reinforce one another, sustaining the community until a point where the feedback loop breaks down, resulting in abrupt collapse. 

\subsection*{Static Models Underestimate Community Fragility}
\begin{figure}[H]
    \centering
    \includegraphics[width=\linewidth]{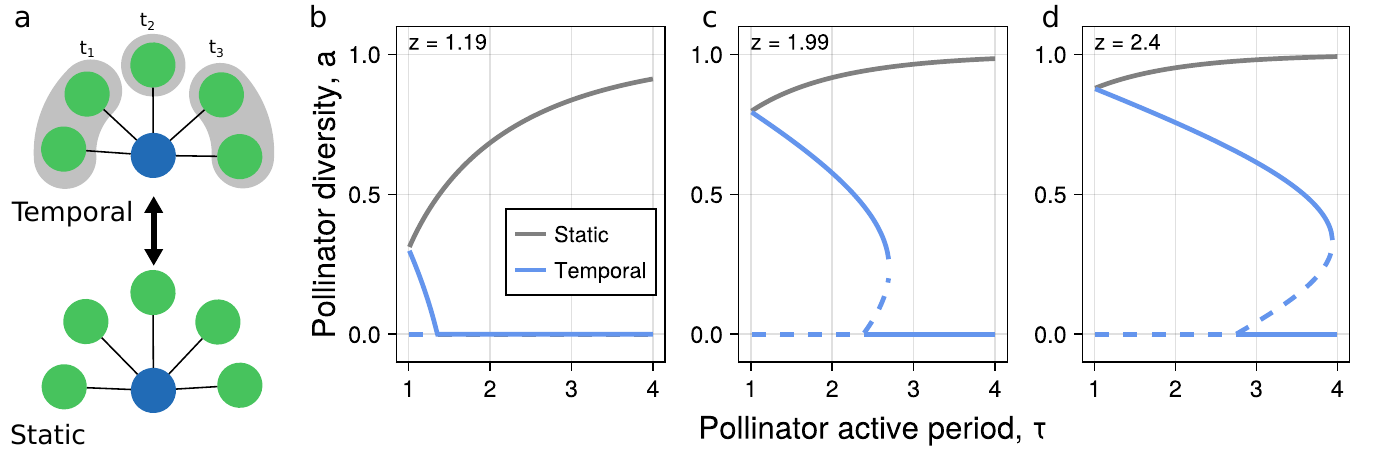}
    \caption{\textbf{Temporal structure reduces community diversity} a) Pollinator persistence with and without temporal structure. In the temporal network, interactions are partitioned by the steps in which they take place (grey partitions). This structure is removed in the static network where all plant neighbours are considered at once. b-d) Plots showing the response of pollinator diversity in the temporal (blue) and static (grey) models across pollinator active period length $\tau$. Panels show a range of mean degree $z$ values.}
    \label{fig:Static_network}
\end{figure}

To illustrate the impact of including temporal conditions for pollinator persistence, we contrast our model with a static, temporally-aggregated baseline. In this case, we ``flatten'' the network, so pollinator nodes have same number of links, but no partitioning of interactions across time. This creates a much simpler rule for persistence: a pollinator is feasible if at least one of its plant neighbours (at any time) is feasible (Fig~\ref{fig:Static_network}a). Applying this results in slight modification to equations~\ref{eq:sys} and \ref{eq:sys_neighbour} which can be solved to find community diversity as before (see Supplementary Material).

Comparing the two models demonstrates how including temporal structure completely inverts the effect of pollinator active period relative to the static baseline (Fig~\ref{fig:Static_network}b-d). In the static model a longer active period $\tau$ always results in higher community diversity because longer active periods increase a pollinator's total degree, making them more connected and thus more likely to have a feasible neighbour. In the temporal model the opposite is true. The stricter requirement for continuous resources means each additional time step is a potential point of failure which might act as a bottleneck for pollinator persistence. This temporal structure not only reduces diversity, but also fundamentally changes the nature of the transition, introducing discontinuous jumps in diversity not seen in the static case. 

\subsection*{Temporal Structure Reduces Network Robustness}
To assess the resilience of the temporal plant–pollinator networks we consider community robustness to random species loss. Specifically, we calculate the proportion of species that persist following the removal of a fraction $r$ of species in the network, accounting for both the loss from the initial removal, and the secondary extinctions that propagate through the community. 

Our generating function framework provides an elegant way to compute the final community state after these types of disturbance by introducing an attack function, $R(x) = (1-r)x + r$, where $r$ is the proportion of nodes removed \parencite{gross_network_2022}. We compose the attack function with our original degree generating functions to account for node removal via its effect on the degree distributions. This results in a set of modified generating functions that describe the system after the attack
\begin{align}
    \tilde{P}(x) = P(R(x)) \quad \tilde{A}(x) = A(R(x))
\end{align}
These modified generating functions can then be used with equations~\ref{eq:sys} and \ref{eq:sys_neighbour} to find the final community state.

\begin{figure}[H]
    \centering
    \includegraphics[width=0.5\linewidth]{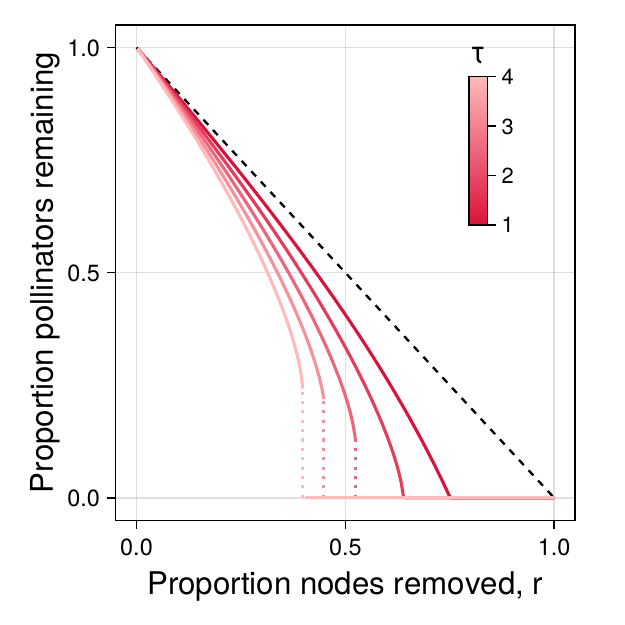}
    \caption{\textbf{Robustness of plant–pollinator networks to node removal} The proportion of original pollinator diversity that remains (y-axis) following the removal of a proportion of species (plants and pollinators) from the community (x-axis). Each line shows the response of a network with varying values of $\tau$ (colours) and a mean degree $z=4.0$. The dashed 1:1 line shows the expectation with no additional secondary extinctions. The loss of species is initially continuous but we see catastrophic disruption once the value of $\tau$ is large enough due to the temporal structure in the network.}
    \label{fig:robustness}
\end{figure}

Our analysis shows how the temporal structure of the community, captured in $\tau$, drives robustness to species loss (Fig~\ref{fig:robustness}). When pollinators persist for only a single time step and networks have no temporal linkages (i.e. $\tau = 1$), communities are maximally robust to attack. As $\tau$ increases, communities become more susceptible to species loss, displaying greater numbers of secondary extinctions for a given attack size $r$. For sufficiently high $\tau$ values the discontinuous transition seen in Fig~\ref{fig:Fig1} manifests, resulting catastrophic collapses before even 50\% of nodes are removed.

\section*{Discussion}
In this paper, we explore the role of temporal structure in determining the diversity and robustness of plant–pollinator communities. Using tools from percolation theory, we demonstrate how the specific requirements for population persistence across time determine the emergent diversity that communities can support. We find that temporal structure fundamentally alters community robustness, increases sensitivity to disturbance, and creates the potential for abrupt, discontinuous transitions. These results highlight how the temporal dependencies inherent in plant–pollinator networks can act as a driver of their robustness, and in some cases, cause catastrophic collapse. 

Our results suggest that resource continuity is a fundamental driver in the emergence of community-scale properties in plant–pollinator networks.  While previous work has established how the distribution of resources across time limits pollinator feasibility \parencite{schellhorn_time_2015, hemberger_floral_2023}, our model demonstrates how these constraints propagate within a larger interaction network. When pollinators require continuous resources to survive, any period of scarcity can act as a temporal bottleneck, potentially triggering population extinction. Crucially, this failure does not occur in isolation; it also removes a functional partner for plants at other time steps, which, in turn, may cause cascades of secondary extinctions. Consequently, the state of a community not only depend on the structure of interactions within any time step, but also on the coupling across the entire season. 

The strength of this temporal coupling translates directly into a reduction in community robustness, potentially triggering complete collapse once interdependencies pass a critical threshold (Fig~\ref{fig:robustness}). This vulnerability is clearly illustrated by the comparison to static baselines; when interactions are aggregated over time, the resulting networks consistently displaying higher diversity and an absence of abrupt transitions (Fig~\ref{fig:Static_network}). While this static case is a simplification, it serves to highlight how failing to account for temporal structure may mask the fragility of these systems. 

These results reinforce the broad consensus regarding the importance of temporal effects in plant–pollinator systems \parencite{caradonna_seeing_2021} and highlight the importance of accounting for temporal structure when assessing robustness. While approaches that rely on comparing robustness between snapshots effectively capture the immediate topology of plant–pollinator networks \parencite{petanidou_long-term_2008, biella_network_2017, souza_temporal_2018}, they may systematically overestimate robustness by overlooking dependencies that span across time. Ultimately because community persistence is dictated by this coupling, robustness can only be accurately evaluated by considering the entire seasonal network as a single, integrated unit.

In our model, we find that system-wide collapses are induced by discontinuous transitions, points where small perturbations can cause sudden, catastrophic jumps in community diversity. This behaviour is driven by the asymmetric requirements of plants and pollinators, where the stricter feasibility conditions for pollinators create a structural bottleneck for the entire community. This finding mirrors previous work where it was shown that similar asymmetries cause analogous transitions in a model of microbial cross-feeding \parencite{clegg_cross-feeding_2025}. The re-emergence of this phenomenon in a different ecological context suggests that this mechanism is not unique to the system we study here, but may reflect a broader ecological principle. Specifically, in systems where persistence is dictated by asymmetric dependences, we may expect the emergence of structural vulnerabilities and the potential for abrupt, discontinuous transitions.  

The ability to isolate these behaviours also highlights the unique power of the structural methods we use to identify the drivers of community collapse. These insights are often lost in traditional dynamical models or computational simulations, where noise and finite-size effects can mask the exact nature of underlying critical transitions. By focusing the structural features of the system, we reveal how the organisation of temporal networks creates specific, structural vulnerabilities. As our results demonstrate, this framework provides a powerful way to assess the resilience of plant–pollinator communities and suggests that such approaches may help explain collapses observed in other types of models \parencite{lever_sudden_2014,bascompte_resilience_2023}, and increase our understanding of the viability of complex ecological systems more generally. 

Finally, the structural vulnerability identified in our model has significant implications for the human-induced pressures faced by plant–pollinator networks in the Anthropocene. Our finding of the importance of temporal bottlenecks suggests that plant–pollinator systems may be particularly vulnerable to drivers that homogenise temporal structure or disrupt phenology. For example, agricultural intensification often results in landscape-level monocultures with brief, synchronised flowering periods \parencite{vasseur_cropping_2013, hemberger_floral_2023}; our model suggests that such temporal structure is unlikely to support diverse plant–pollinator communities, rendering those that persist fragile to perturbation. Similarly, climate-induced phenological shifts can decouple plant and pollinator life-history events \parencite{memmott_global_2007,kudo_when_2019, revilla_robustness_2015}. In our framework, these shifts cause temporal mismatches which effectively cause reductions in connectivity, increasing the potential for bottlenecks that may push communities to low diversity states or precipitate collapse. Our findings thus underscore that maintaining the temporal continuity of resources is as vital for community persistence.

\printbibliography

@article{bascompte_resilience_2023,
	title = {The {Resilience} of {Plant}–{Pollinator} {Networks}},
	volume = {68},
	issn = {0066-4170, 1545-4487},
	url = {https://www.annualreviews.org/content/journals/10.1146/annurev-ento-120120-102424},
	doi = {10.1146/annurev-ento-120120-102424},
	language = {en},
	number = {Volume 68, 2023},
	urldate = {2026-04-08},
	journal = {Annual Review of Entomology},
	publisher = {Annual Reviews},
	author = {Bascompte, Jordi and Scheffer, Marten},
	month = jan,
	year = {2023},
	pages = {363--380},
}

@article{erdos_random_1959,
	title = {On random graphs. {I}.},
	volume = {6},
	issn = {00333883},
	url = {https://publi.math.unideb.hu/load_doi.php?pdoi=10_5486_PMD_1959_6_3_4_12},
	number = {3-4},
	urldate = {2026-04-07},
	journal = {Publicationes Mathematicae Debrecen},
	author = {Erdős, P. and Rényi, A.},
	year = {1959},
	pages = {290--297},
}

@article{knight_pollen_2005,
	title = {Pollen {Limitation} of {Plant} {Reproduction}: {Pattern} and {Process}},
	volume = {36},
	issn = {1543-592X, 1545-2069},
	shorttitle = {Pollen {Limitation} of {Plant} {Reproduction}},
	url = {https://www.annualreviews.org/doi/10.1146/annurev.ecolsys.36.102403.115320},
	doi = {10.1146/annurev.ecolsys.36.102403.115320},
	language = {en},
	number = {1},
	urldate = {2026-04-07},
	journal = {Annual Review of Ecology, Evolution, and Systematics},
	author = {Knight, Tiffany M. and Steets, Janette A. and Vamosi, Jana C. and Mazer, Susan J. and Burd, Martin and Campbell, Diane R. and Dudash, Michele R. and Johnston, Mark O. and Mitchell, Randall J. and Ashman, Tia-Lynn},
	month = dec,
	year = {2005},
	pages = {467--497},
}

@article{timberlake_network_2022,
	title = {A network approach for managing ecosystem services and improving food and nutrition security on smallholder farms},
	volume = {4},
	copyright = {© 2022 The Authors. People and Nature published by John Wiley \& Sons Ltd on behalf of British Ecological Society},
	issn = {2575-8314},
	url = {https://onlinelibrary.wiley.com/doi/abs/10.1002/pan3.10295},
	doi = {10.1002/pan3.10295},
	language = {en},
	number = {2},
	urldate = {2026-04-07},
	journal = {People and Nature},
	author = {Timberlake, Thomas P. and Cirtwill, Alyssa R. and Baral, Sushil C. and Bhusal, Daya R. and Devkota, Kedar and Harris-Fry, Helen A. and Kortsch, Susanne and Myers, Samuel S. and Roslin, Tomas and Saville, Naomi M. and Smith, Matthew R. and Strona, Giovanni and Memmott, Jane},
	year = {2022},
	note = {\_eprint: https://besjournals.onlinelibrary.wiley.com/doi/pdf/10.1002/pan3.10295},
	pages = {563--575},
}

@article{bastolla_architecture_2009,
	title = {The architecture of mutualistic networks minimizes competition and increases biodiversity},
	volume = {458},
	copyright = {2009 Macmillan Publishers Limited. All rights reserved},
	issn = {1476-4687},
	url = {https://www.nature.com/articles/nature07950},
	doi = {10.1038/nature07950},
	language = {en},
	number = {7241},
	urldate = {2026-04-07},
	journal = {Nature},
	publisher = {Nature Publishing Group},
	author = {Bastolla, Ugo and Fortuna, Miguel A. and Pascual-García, Alberto and Ferrera, Antonio and Luque, Bartolo and Bascompte, Jordi},
	month = apr,
	year = {2009},
	pages = {1018--1020},
}

@article{lever_sudden_2014,
	title = {The sudden collapse of pollinator communities},
	volume = {17},
	copyright = {© 2014 John Wiley \& Sons Ltd/CNRS},
	issn = {1461-0248},
	url = {https://onlinelibrary.wiley.com/doi/abs/10.1111/ele.12236},
	doi = {10.1111/ele.12236},
	language = {en},
	number = {3},
	urldate = {2026-03-09},
	journal = {Ecology Letters},
	author = {Lever, J. Jelle and van Nes, Egbert H. and Scheffer, Marten and Bascompte, Jordi},
	year = {2014},
	note = {\_eprint: https://onlinelibrary.wiley.com/doi/pdf/10.1111/ele.12236},
	pages = {350--359},
}

@article{clegg_cross-feeding_2025,
	title = {Cross-feeding creates tipping points in microbiome diversity},
	volume = {122},
	url = {https://www.pnas.org/doi/abs/10.1073/pnas.2425603122},
	doi = {10.1073/pnas.2425603122},
	language = {en},
	number = {19},
	urldate = {2026-03-02},
	journal = {Proceedings of the National Academy of Sciences},
	publisher = {Proceedings of the National Academy of Sciences},
	author = {Clegg, Tom and Gross, Thilo},
	month = may,
	year = {2025},
	pages = {e2425603122},
}

@article{memmott_global_2007,
	title = {Global warming and the disruption of plant–pollinator interactions},
	volume = {10},
	issn = {1461-0248},
	url = {https://onlinelibrary.wiley.com/doi/abs/10.1111/j.1461-0248.2007.01061.x},
	doi = {10.1111/j.1461-0248.2007.01061.x},
	language = {en},
	number = {8},
	urldate = {2026-02-10},
	journal = {Ecology Letters},
	author = {Memmott, Jane and Craze, Paul G. and Waser, Nickolas M. and Price, Mary V.},
	year = {2007},
	note = {\_eprint: https://onlinelibrary.wiley.com/doi/pdf/10.1111/j.1461-0248.2007.01061.x},
	pages = {710--717},
}

@article{olesen_missing_2010,
	title = {Missing and forbidden links in mutualistic networks},
	volume = {278},
	issn = {0962-8452},
	url = {https://doi.org/10.1098/rspb.2010.1371},
	doi = {10.1098/rspb.2010.1371},
	number = {1706},
	urldate = {2026-02-10},
	journal = {Proceedings of the Royal Society B: Biological Sciences},
	author = {Olesen, Jens M. and Bascompte, Jordi and Dupont, Yoko L. and Elberling, Heidi and Rasmussen, Claus and Jordano, Pedro},
	month = sep,
	year = {2010},
	pages = {725--732},
}

@article{rathcke_phenological_1985,
	title = {Phenological {Patterns} of {Terrestrial} {Plants}},
	volume = {16},
	issn = {1543-592X, 1545-2069},
	url = {https://www.annualreviews.org/content/journals/10.1146/annurev.es.16.110185.001143},
	doi = {10.1146/annurev.es.16.110185.001143},
	language = {en},
	number = {Volume 16, 1985},
	urldate = {2026-02-10},
	journal = {Annual Review of Ecology, Evolution, and Systematics},
	publisher = {Annual Reviews},
	author = {Rathcke, B. and Lacey, E. P.},
	month = nov,
	year = {1985},
	pages = {179--214},
}

@article{memmott_tolerance_2004,
	title = {Tolerance of pollination networks to species extinctions},
	volume = {271},
	issn = {0962-8452},
	url = {https://doi.org/10.1098/rspb.2004.2909},
	doi = {10.1098/rspb.2004.2909},
	number = {1557},
	urldate = {2026-02-10},
	journal = {Proceedings of the Royal Society B: Biological Sciences},
	author = {Memmott, Jane and Waser, Nickolas M. and Price, Mary V.},
	month = dec,
	year = {2004},
	pages = {2605--2611},
}

@article{bascompte_plant-animal_2007,
	title = {Plant-{Animal} {Mutualistic} {Networks}: {The} {Architecture} of {Biodiversity}},
	volume = {38},
	issn = {1543-592X, 1545-2069},
	shorttitle = {Plant-{Animal} {Mutualistic} {Networks}},
	url = {https://www.annualreviews.org/content/journals/10.1146/annurev.ecolsys.38.091206.095818},
	doi = {10.1146/annurev.ecolsys.38.091206.095818},
	language = {en},
	number = {Volume 38, 2007},
	urldate = {2026-02-10},
	journal = {Annual Review of Ecology, Evolution, and Systematics},
	publisher = {Annual Reviews},
	author = {Bascompte, Jordi and Jordano, Pedro},
	month = dec,
	year = {2007},
	pages = {567--593},
}

@article{revilla_robustness_2015,
	title = {Robustness of mutualistic networks under phenological change and habitat destruction},
	volume = {124},
	copyright = {© 2014 The Authors},
	issn = {1600-0706},
	url = {https://onlinelibrary.wiley.com/doi/abs/10.1111/oik.01532},
	doi = {10.1111/oik.01532},
	language = {en},
	number = {1},
	urldate = {2025-11-26},
	journal = {Oikos},
	author = {Revilla, Tomás A. and Encinas-Viso, Francisco and Loreau, Michel},
	year = {2015},
	note = {\_eprint: https://nsojournals.onlinelibrary.wiley.com/doi/pdf/10.1111/oik.01532},
	pages = {22--32},
}

@article{trojelsgaard_ecological_2016,
	title = {Ecological networks in motion: micro- and macroscopic variability across scales},
	volume = {30},
	copyright = {© 2016 The Authors. Functional Ecology © 2016 British Ecological Society},
	issn = {1365-2435},
	shorttitle = {Ecological networks in motion},
	url = {https://onlinelibrary.wiley.com/doi/abs/10.1111/1365-2435.12710},
	doi = {10.1111/1365-2435.12710},
	language = {en},
	number = {12},
	urldate = {2025-11-26},
	journal = {Functional Ecology},
	author = {Trøjelsgaard, Kristian and Olesen, Jens M.},
	year = {2016},
	note = {\_eprint: https://besjournals.onlinelibrary.wiley.com/doi/pdf/10.1111/1365-2435.12710},
	pages = {1926--1935},
}

@article{nicholson_corridors_2021,
	title = {Corridors through time: {Does} resource continuity impact pollinator communities, populations, and individuals?},
	volume = {31},
	copyright = {© 2020 by the Ecological Society of America},
	issn = {1939-5582},
	shorttitle = {Corridors through time},
	url = {https://onlinelibrary.wiley.com/doi/abs/10.1002/eap.2260},
	doi = {10.1002/eap.2260},
	language = {en},
	number = {3},
	urldate = {2025-11-26},
	journal = {Ecological Applications},
	author = {Nicholson, Charlie C. and J.-M. Hayes, Jen and Connolly, Samantha and Ricketts, Taylor H.},
	year = {2021},
	note = {\_eprint: https://esajournals.onlinelibrary.wiley.com/doi/pdf/10.1002/eap.2260},
	pages = {e02260},
}

@article{kudo_when_2019,
	title = {When spring ephemerals fail to meet pollinators: mechanism of phenological mismatch and its impact on plant reproduction},
	copyright = {© 2019 The Author(s)},
	shorttitle = {When spring ephemerals fail to meet pollinators},
	url = {https://royalsocietypublishing.org/doi/10.1098/rspb.2019.0573},
	doi = {10.1098/rspb.2019.0573},
	language = {EN},
	urldate = {2025-11-26},
	journal = {Proceedings of the Royal Society B},
	publisher = {The Royal Society},
	author = {Kudo, Gaku and Cooper, Elisabeth J.},
	month = jun,
	year = {2019},
}

@article{russo_supporting_2013,
	title = {Supporting crop pollinators with floral resources: network-based phenological matching},
	volume = {3},
	copyright = {© 2013 The Authors. Ecology and Evolution published by John Wiley \& Sons Ltd.},
	issn = {2045-7758},
	shorttitle = {Supporting crop pollinators with floral resources},
	url = {https://onlinelibrary.wiley.com/doi/abs/10.1002/ece3.703},
	doi = {10.1002/ece3.703},
	language = {en},
	number = {9},
	urldate = {2025-11-26},
	journal = {Ecology and Evolution},
	author = {Russo, Laura and DeBarros, Nelson and Yang, Suann and Shea, Katriona and Mortensen, David},
	year = {2013},
	note = {\_eprint: https://onlinelibrary.wiley.com/doi/pdf/10.1002/ece3.703},
	pages = {3125--3140},
}

@article{souza_temporal_2018,
	title = {Temporal variation in plant–pollinator networks from seasonal tropical environments: {Higher} specialization when resources are scarce},
	volume = {106},
	issn = {1365-2745},
	shorttitle = {Temporal variation in plant–pollinator networks from seasonal tropical environments},
	url = {https://onlinelibrary.wiley.com/doi/abs/10.1111/1365-2745.12978},
	doi = {10.1111/1365-2745.12978},
	language = {en},
	number = {6},
	urldate = {2025-11-26},
	journal = {Journal of Ecology},
	author = {Souza, Camila S. and Maruyama, Pietro K. and Aoki, Camila and Sigrist, Maria R. and Raizer, Josué and Gross, Caroline L. and de Araujo, Andréa C.},
	year = {2018},
	note = {\_eprint: https://besjournals.onlinelibrary.wiley.com/doi/pdf/10.1111/1365-2745.12978},
	pages = {2409--2420},
}

@article{ogilvie_interactions_2017,
	series = {Pests and resistance * {Behavioural} ecology},
	title = {Interactions between bee foraging and floral resource phenology shape bee populations and communities},
	volume = {21},
	issn = {2214-5745},
	url = {https://www.sciencedirect.com/science/article/pii/S2214574517300408},
	doi = {10.1016/j.cois.2017.05.015},
	urldate = {2025-11-26},
	journal = {Current Opinion in Insect Science},
	author = {Ogilvie, Jane E and Forrest, Jessica RK},
	month = jun,
	year = {2017},
	pages = {75--82},
}

@article{vasseur_cropping_2013,
	series = {Landscape ecology and biodiversity in agricultural landscapes},
	title = {The cropping systems mosaic: {How} does the hidden heterogeneity of agricultural landscapes drive arthropod populations?},
	volume = {166},
	issn = {0167-8809},
	shorttitle = {The cropping systems mosaic},
	url = {https://www.sciencedirect.com/science/article/pii/S0167880912003246},
	doi = {10.1016/j.agee.2012.08.013},
	urldate = {2025-11-26},
	journal = {Agriculture, Ecosystems \& Environment},
	author = {Vasseur, Chloé and Joannon, Alexandre and Aviron, Stéphanie and Burel, Françoise and Meynard, Jean-Marc and Baudry, Jacques},
	month = feb,
	year = {2013},
	pages = {3--14},
}

@article{schellhorn_time_2015,
	title = {Time will tell: resource continuity bolsters ecosystem services},
	volume = {30},
	issn = {0169-5347},
	shorttitle = {Time will tell},
	url = {https://www.sciencedirect.com/science/article/pii/S0169534715001585},
	doi = {10.1016/j.tree.2015.06.007},
	number = {9},
	urldate = {2025-11-26},
	journal = {Trends in Ecology \& Evolution},
	author = {Schellhorn, Nancy A and Gagic, Vesna and Bommarco, Riccardo},
	month = sep,
	year = {2015},
	pages = {524--530},
}

@article{hemberger_floral_2023,
	title = {Floral resource discontinuity contributes to spatial mismatch between pollinator supply and pollination demand in a pollinator-dependent agricultural landscapes},
	volume = {38},
	issn = {1572-9761},
	url = {https://doi.org/10.1007/s10980-023-01707-w},
	doi = {10.1007/s10980-023-01707-w},
	language = {en},
	number = {12},
	urldate = {2025-11-26},
	journal = {Landscape Ecology},
	author = {Hemberger, Jeremy and Gratton, Claudio},
	month = dec,
	year = {2023},
	pages = {4439--4450},
}

@article{lowe_temporal_2023,
	title = {Temporal change in floral availability leads to periods of resource limitation and affects diet specificity in a generalist pollinator},
	volume = {32},
	issn = {0962-1083},
	url = {https://onlinelibrary.wiley.com/doi/full/10.1111/mec.16719},
	doi = {10.1111/mec.16719},
	number = {23},
	urldate = {2025-11-26},
	journal = {Molecular Ecology},
	publisher = {John Wiley \& Sons, Ltd},
	author = {Lowe, Abigail and Jones, Laura and Brennan, Georgina and Creer, Simon and Christie, Lynda and de Vere, Natasha},
	month = dec,
	year = {2023},
	pages = {6363--6376},
}

@article{biella_network_2017,
	title = {Network analysis of phenological units to detect important species in plant-pollinator assemblages: can it inform conservation strategies?},
	volume = {18},
	issn = {1588-2756},
	shorttitle = {Network analysis of phenological units to detect important species in plant-pollinator assemblages},
	url = {https://doi.org/10.1556/168.2017.18.1.1},
	doi = {10.1556/168.2017.18.1.1},
	language = {en},
	number = {1},
	urldate = {2025-11-14},
	journal = {Community Ecology},
	author = {Biella, P. and Ollerton, J. and Barcella, M. and Assini, S.},
	month = apr,
	year = {2017},
	pages = {1--10},
}

@article{caradonna_interaction_2017,
	title = {Interaction rewiring and the rapid turnover of plant–pollinator networks},
	volume = {20},
	copyright = {© 2017 John Wiley \& Sons Ltd/CNRS},
	issn = {1461-0248},
	url = {https://onlinelibrary.wiley.com/doi/abs/10.1111/ele.12740},
	doi = {10.1111/ele.12740},
	language = {en},
	number = {3},
	urldate = {2025-10-21},
	journal = {Ecology Letters},
	author = {CaraDonna, Paul J. and Petry, William K. and Brennan, Ross M. and Cunningham, James L. and Bronstein, Judith L. and Waser, Nickolas M. and Sanders, Nathan J.},
	year = {2017},
	note = {\_eprint: https://onlinelibrary.wiley.com/doi/pdf/10.1111/ele.12740},
	pages = {385--394},
}

@article{petanidou_long-term_2008,
	title = {Long-term observation of a pollination network: fluctuation in species and interactions, relative invariance of network structure and implications for estimates of specialization},
	volume = {11},
	copyright = {© 2008 Blackwell Publishing Ltd/CNRS},
	issn = {1461-0248},
	shorttitle = {Long-term observation of a pollination network},
	url = {https://onlinelibrary.wiley.com/doi/abs/10.1111/j.1461-0248.2008.01170.x},
	doi = {10.1111/j.1461-0248.2008.01170.x},
	language = {en},
	number = {6},
	urldate = {2025-10-20},
	journal = {Ecology Letters},
	author = {Petanidou, Theodora and Kallimanis, Athanasios S. and Tzanopoulos, Joseph and Sgardelis, Stefanos P. and Pantis, John D.},
	year = {2008},
	note = {\_eprint: https://onlinelibrary.wiley.com/doi/pdf/10.1111/j.1461-0248.2008.01170.x},
	pages = {564--575},
}

@article{caradonna_seeing_2021,
	title = {Seeing through the static: the temporal dimension of plant–animal mutualistic interactions},
	volume = {24},
	copyright = {© 2020 The Authors. Ecology Letters published by John Wiley \& Sons Ltd.},
	issn = {1461-0248},
	shorttitle = {Seeing through the static},
	url = {https://onlinelibrary.wiley.com/doi/abs/10.1111/ele.13623},
	doi = {10.1111/ele.13623},
	language = {en},
	number = {1},
	urldate = {2025-10-20},
	journal = {Ecology Letters},
	author = {CaraDonna, Paul J. and Burkle, Laura A. and Schwarz, Benjamin and Resasco, Julian and Knight, Tiffany M. and Benadi, Gita and Blüthgen, Nico and Dormann, Carsten F. and Fang, Qiang and Fründ, Jochen and Gauzens, Benoit and Kaiser-Bunbury, Christopher N. and Winfree, Rachael and Vázquez, Diego P.},
	year = {2021},
	note = {\_eprint: https://onlinelibrary.wiley.com/doi/pdf/10.1111/ele.13623},
	pages = {149--161},
}

@article{caradonna_temporal_2020,
	title = {Temporal flexibility in the structure of plant–pollinator interaction networks},
	volume = {129},
	copyright = {© 2020 Nordic Society Oikos. Published by John Wiley \& Sons Ltd},
	issn = {1600-0706},
	url = {https://onlinelibrary.wiley.com/doi/abs/10.1111/oik.07526},
	doi = {10.1111/oik.07526},
	language = {en},
	number = {9},
	urldate = {2025-10-20},
	journal = {Oikos},
	author = {CaraDonna, Paul J. and Waser, Nickolas M.},
	year = {2020},
	note = {\_eprint: https://nsojournals.onlinelibrary.wiley.com/doi/pdf/10.1111/oik.07526},
	pages = {1369--1380},
}

@article{lampo_structural_2024,
	title = {Structural dynamics of plant–pollinator mutualistic networks},
	volume = {3},
	issn = {2752-6542},
	url = {https://doi.org/10.1093/pnasnexus/pgae209},
	doi = {10.1093/pnasnexus/pgae209},
	number = {6},
	urldate = {2025-10-20},
	journal = {PNAS Nexus},
	author = {Lampo, Aniello and Palazzi, María J and Borge-Holthoefer, Javier and Solé-Ribalta, Albert},
	month = jun,
	year = {2024},
	pages = {pgae209},
}

@article{molloy_critical_1995,
	title = {A critical point for random graphs with a given degree sequence},
	volume = {6},
	issn = {1098-2418},
	url = {https://onlinelibrary.wiley.com/doi/full/10.1002/rsa.3240060204},
	doi = {10.1002/RSA.3240060204},
	number = {2-3},
	urldate = {2024-11-12},
	journal = {Random Structures \& Algorithms},
	publisher = {John Wiley \& Sons, Ltd},
	author = {Molloy, Michael and Reed, Bruce},
	month = mar,
	year = {1995},
	pages = {161--180},
}

@article{buldyrev_catastrophic_2010,
	title = {Catastrophic cascade of failures in interdependent networks},
	volume = {464},
	issn = {1476-4687},
	url = {https://www.nature.com/articles/nature08932},
	doi = {10.1038/nature08932},
	number = {7291},
	urldate = {2024-11-12},
	journal = {Nature 2010 464:7291},
	publisher = {Nature Publishing Group},
	author = {Buldyrev, Sergey V. and Parshani, Roni and Paul, Gerald and Stanley, H. Eugene and Havlin, Shlomo},
	month = apr,
	year = {2010},
	note = {arXiv: 0907.1182},
	pages = {1025--1028},
}

@article{gross_network_2022,
	title = {Network {Robustness} {Revisited}},
	volume = {10},
	issn = {2296424X},
	url = {www.frontiersin.org},
	doi = {10.3389/FPHY.2022.823564/BIBTEX},
	urldate = {2024-09-10},
	journal = {Frontiers in Physics},
	publisher = {Frontiers Media SA},
	author = {Gross, Thilo and Barth, Laura},
	month = jun,
	year = {2022},
	note = {arXiv: 2202.07911},
	pages = {823564},
}

@article{newman_random_2001,
	title = {Random graphs with arbitrary degree distributions and their applications},
	volume = {64},
	issn = {1063651X},
	url = {https://journals.aps.org/pre/abstract/10.1103/PhysRevE.64.026118},
	doi = {10.1103/PhysRevE.64.026118},
	number = {2},
	urldate = {2024-09-10},
	journal = {Physical Review E},
	publisher = {American Physical Society},
	author = {Newman, M. E.J. and Strogatz, S. H. and Watts, D. J.},
	month = jul,
	year = {2001},
	note = {arXiv: cond-mat/0007235},
	pages = {026118},
}

@article{callaway_network_2000,
	title = {Network {Robustness} and {Fragility}: {Percolation} on {Random} {Graphs}},
	volume = {85},
	issn = {00319007},
	url = {https://journals.aps.org/prl/abstract/10.1103/PhysRevLett.85.5468},
	doi = {10.1103/PhysRevLett.85.5468},
	number = {25},
	urldate = {2024-09-02},
	journal = {Physical Review Letters},
	publisher = {American Physical Society},
	author = {Callaway, Duncan S. and Newman, M. E.J. and Strogatz, Steven H. and Watts, Duncan J.},
	month = dec,
	year = {2000},
	note = {arXiv: cond-mat/0007300},
	pages = {5468},
}

\end{document}